\documentstyle[preprint,epsf,aps]{revtex}
\renewcommand{\v}[1]{{\mathbf{#1}}} 
 
\renewcommand{\sp}[2]{#1\! \cdot \!  #2} 
 
\newcommand{\s}[1]{\mbox{sign}(#1)} 
 
\renewcommand{\d}[1]{d#1} 
\newcommand{\dd}[2] {\frac{d#1}{d#2}} 

\newcommand{\erf}{\mbox{erf}}

\newcommand{\vw }[1]{\v{w} _{#1}} 
\newcommand{\vwp}[1]{\v{w} _{#1}^{+}} 
\newcommand{\w }[1]{w_{#1}}

\renewcommand{\c }{\cos  (\theta) } 
\newcommand{\x }{\v{x}} 
\renewcommand{\a }{\alpha} 
\newcommand{\C }{\v{C}}  
\newcommand{\vg}[1]{\v{g}_{#1}}

\newcommand{\bea}{\begin{eqnarray}} 
\newcommand{\eea}{\end{eqnarray}} 
\newcommand{\be}{\begin{equation}} 
\newcommand{\ee}{\end{equation}} 
\newcommand{\sm}[1]{\sigma_{#1}}

\begin{document}  
\title{Interacting Neural Networks}   
\author{W. Kinzel and R. Metzler}
\address{Institut f\"{u}r Theoretische Physik, 
Universit\"{a}t W\"{u}rzburg, Am Hubland, D-97074 W\"{u}rzburg, Germany}
\author{I. Kanter}
\address{Minerva Center and Department of Physics, Bar Ilan University, 52900 
Ramat Gan, Israel}  
\date{\today}
\maketitle

\begin{abstract}
Several scenarios of interacting neural networks which are trained
either in an identical or in a competitive way are solved 
analytically. In the case of identical training each perceptron 
receives the output of its neighbour. The symmetry of the stationary 
state as well as the sensitivity to the used training algorithm 
are investigated. 
Two competitive perceptrons trained on mutually exclusive 
learning aims and a perceptron which is trained on the opposite
of its own output are examined analytically. 
An ensemble of competitive perceptrons is
used as decision-making algorithms in 
a model of a closed market (El Farol Bar problem or Minority Game);
each network is trained on the history of minority decisions.
This ensemble of perceptrons relaxes to a stationary state 
whose performance can be better than random.

\end{abstract}
Simple models of neural networks describe a wide variety of phenomena in
neurobiology and information theory. Neural networks are systems of
elements interacting by adaptive couplings which are trained by a set of
examples. After training they function as content addressable
associative memory, as classifiers or as prediction algorithms. Using
methods of statistical physics many of these phenomena have been
elucidated analytically for infinitely large neural networks
\cite{Hertz:NeuralComp,Opper:Generalization}.

Most studies of feed-forward neural networks have
concentrated on a single network learning a fixed rule, 
which is usually 
a second network, the so-called teacher. The teacher
network is presenting examples, sets of input/output
data, and the student network is adapting its weights to 
this set of examples. In an on-line training scenario each 
example is presented only once, hence training is a 
dynamical process \cite{Saad:Online,Biehl:Online}.
The teacher network may also generate a time series of 
output numbers \cite{Eisenstein:BG,Schroeder:Cycles},
and the student learns by following the time series. 
The weights of the teacher network are fixed in this scenario.

Many phenomena in biology, social and computer science may 
be modeled by a system of interacting adaptive algorithms
(see e.g. \cite{Wolpert:Intro}). However, little is known about general 
properties of such systems. In this paper we derive an analytic
solution of a system of interacting neural networks. Each network 
is a simple perceptron with an $N$-dimensional weight vector.
These networks receive an identical input vector, produce
output bits and learn from each other. In Section \ref{SEC-Mutual},
each network is trained by the output of its neighbour, with a 
cyclic flow of information. By iterating the training step for 
randomly chosen input vectors, the dynamical process relaxes 
to a stationary state. In the limit of $N\rightarrow \infty$
we describe the process by ordinary differential equations
for a few order parameters, similiar to the usual student/teacher
scenario \cite{Saad:Online,Biehl:Drift}. We identify the symmetries 
of the stationary state and find phase transitions when increasing the 
learning rate of the training steps.

In Section \ref{SEC-Compet} and \ref{SEC-Conteach} 
we study different training scenarios 
with two interacting perceptrons and various learning 
algorithms.

In Section \ref{SEC-Minority} we apply the system of interacting networks
to a problem of game theory called the minority game, 
which is derived from the El-Farol Bar problem 
\cite{Arthur:ElFarol,Challet:Emerg.}.
We consider a set of agents who have to make a binary decision.
Each agent wins only if he/she belongs to the minority of all
decisions. This process is iterated. Each agent has to develop
an algorithm which makes a decision according to the history 
of the global minority decisions. The problem
recently received a lot of attention in the context of statistical 
physics \cite{Minority}. Here we follow a novel approach:
Each agent uses a perceptron for making his/her decision,
and each perceptron is trained on the minority of all output bits.

\section{Mutual Learning, symmetric case}
\label{SEC-Mutual}
In this section we investigate a system of interacting neural
networks as follows: several identical networks are arranged on
an oriented ring. All networks receive an identical input 
and produce different output according to their weight vectors.
Each network is trained by the output of its neighbour on the
ring. This process is iterated until a stationary state is
reached in which the norms and angles between the weight 
vectors no longer change. We are interested in the properties 
of this stationary state.

We consider the simplest feed-forward networks, an ensemble 
of $K$ simple perceptrons, which are represented by $N$-dimensional
weight vectors $\vw{i}$  $(i=1,\dots,K)$ and which map a common 
input vector $\x$ onto binary outputs $\sm{i} = \s{\sp{\x}{\vw{i}}}$.
As order parameters we use the norms $\w{i}=| \vw{i} |$ and the 
respective overlaps $R_{ij} = \sp{\vw{i}}{\vw{j}}$ or 
$\cos (\theta_{ij}) = \sp{\vw{i}}{\vw{j}} / w_{i} w_{j}$. 
When only two perceptrons are considered, the
subscript is dropped: $\c = \sp{\vw{1}}{\vw{2}}/ w_{1} w_{2}$.
The components of the input vector (or pattern) are 
Gaussian with mean 0 and variance 1, yielding $\sp{\x}{\x} =O(N)$.

The updates are of the form
\be 
\vwp{i} = \vw{i} + (\eta_i/N) f(\sm{i}, s) s\, \x
\ee
for unnormalized weights or
\be 
\vwp{i} = \frac{\vw{i} + (\eta_i/N) f(\sm{i}, s) s\, \x}
           {|\vw{i} + (\eta_i/N) f(\sm{i}, s) s\, \x|}
\ee      
for normalized $\vw{i}$. The $+$ denotes a quantity after
one learning step, $\eta_i$ is the learning rate, $s$ is the
desired output, and $f(\sm{i},s)$, the so-called weight
function, defines the learning algorithm. We mostly use 
$f =1$ (the Hebbian rule, called $H$ from now on)
and $f=\Theta(-\sm{i} s)$ (the perceptron learning rule, 
abbreviated $P$ \cite{Hertz:NeuralComp}), and the 
respective variations where the $\vw{i}$ are kept normalized,
denoted as $HN$ and $PN$ respectively. 

We derive differential equations for the order parameters  
in the thermodynamic limit $N \rightarrow \infty$
by taking the scalar product of the 
update rules and introducing a time variable $\a = p/N$,
where $p$ is the number of patterns shown so far.
We use the analytic tools which were previously developed for the
teacher/student scenario \cite{Biehl:Online,Saad:Online}.
If the order parameters are self-averaging (see \cite{Reents:Self-Avg.} for
criteria of self-averaging in this context), integrating over
the distribution of patterns gives deterministic differential 
equations for the order parameters as $N \rightarrow \infty$. 
The required averages are listed in the appendix.

\subsection{Perceptron learning rule}
\label{GSL-P}  
We first restrict ourselves to two perceptrons that try to come to 
an agreement by learning the output of the respective other
perceptron.

\begin{figure}
  \epsfxsize= 0.65\textwidth
  \epsffile{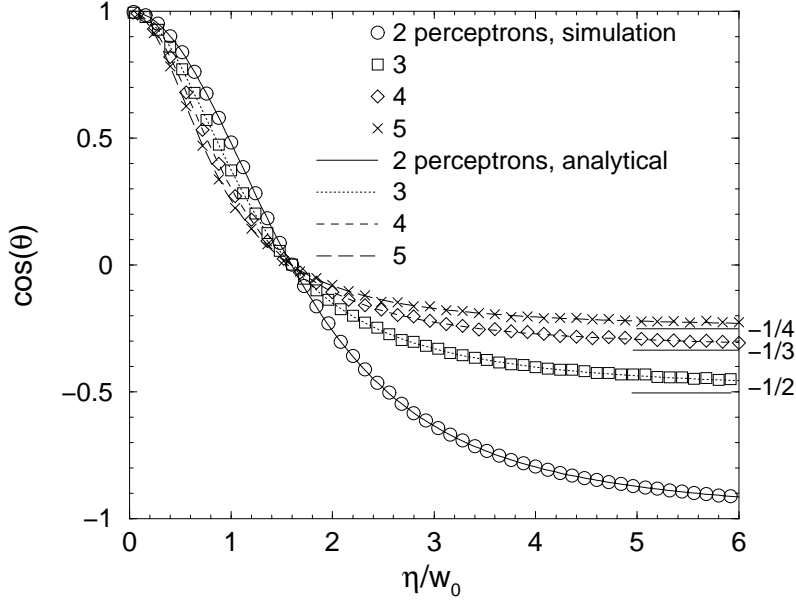}
  \caption{Mutual learning with rule $P$ (see sections \ref{GSL-P} and
    \ref{GSL-K}): comparison between Eqs. (\ref{GSL-fix}) and 
    (\ref{GSLK-fix}) and the stationary state in simulations with 
    $N=100$ and $\a>75$.} 
  \label{GSLP-fig}
\end{figure}

For rule $P$ with identical learning rates $\eta_1=\eta_2=\eta$,
the update  rule is
\bea
\vwp{1} &=& \vw{1} + \frac{\eta}{N} \x\, \sm{2} 
     \Theta(-\sm{1}\,\sm{2}); \nonumber \\
\vwp{2} &=& \vw{2} + \frac{\eta}{N} \x\,\sm{1} \Theta(-\sm{1}\,\sm{2}).
\eea
The sum of both vectors is conserved under this rule: if a 
learning step takes place, it has the same direction and
absolute value, but different signs for the two vectors.  
This conservation can be used to link $\w{1}$ and $\w{2}$ to
$\c$: assuming that  $\w{1} = \w{2} = w$ and starting from 
$\theta_0 = \pi/2$, simple geometry gives $\w{0}/\sqrt{2} 
= \cos (\theta/2) w$. The conservation is also visible in the
differential equations that can be derived using the 
described formalism:
\bea
\dd{\w{1}}{\a} &=& - \frac{\eta}{\sqrt{2 \pi}}(1-\c) + 
\frac{\eta^2 \theta}{2 \w{1}\pi}; \label{GSLP-DGL1} \\
\dd{\w{2}}{\a} &=& - \frac{\eta}{\sqrt{2 \pi}}(1-\c) + 
\frac{\eta^2 \theta}{2 \w{2}\pi}; \label{GSLP-DGL2} \\
\dd{R}{\a} &=& \frac{\eta}{\sqrt{2\pi}} (1-\c)(\w{1}+\w{2})
  - \eta^2 \frac{\theta}{\pi}.\label{GSLP-DGL3}
\eea
If the right-hand side of \ref{GSLP-DGL1} and \ref{GSLP-DGL2}
vanish, so does \ref{GSLP-DGL3}. There is a curve of
fixed points of the system given by the equation
\be
w = \frac{\eta}{\sqrt{2 \pi}} \frac{\theta}{1-\c}.
\label{GSL-fix}
\ee
Using the relation $w = \w{0}/(\sqrt{2} \cos (\theta/2))$,
this can be solved numerically to give the fixed point of
$\c$ as a function of the scaled learning rate $\eta/\w{0}$, 
as shown in Fig. \ref{GSLP-fig}.
 For small learning rates, the perceptrons
come to good agreement, while large $\eta$ leads to antiparallel
vectors.

\begin{figure}
 \epsfxsize= 0.65\textwidth
  \epsffile{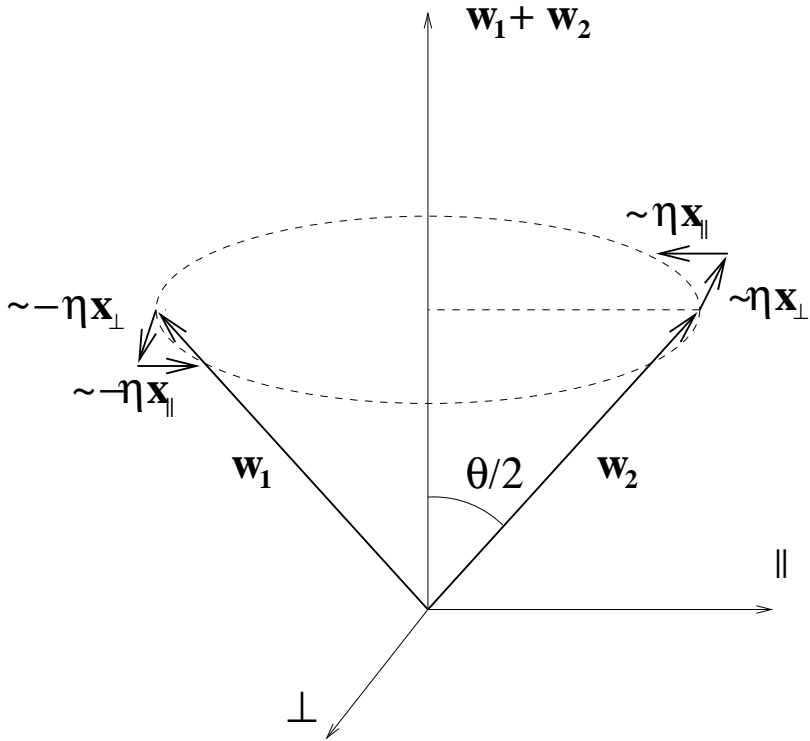}
\caption{Mutual learning with rule $P$: sketch of the geometrical
interpretation. See section \ref{GSL-P}.
 } 
\label{GSL-skizz}
\end{figure}

Geometrically, this can be understood as follows:
each learning step has a component parallel to the plane
spanned by $\vw{1}$ and $\vw{2}$, which decreases the
distance between the vectors, and a perpendicular component,
which increases the distance (see Fig. \ref{GSL-skizz}). 
Equilibrium is reached when a 
typical learning step no longer changes the angle, i.e.
the vectors stay on a cone around $\vw{1}+\vw{2}$. The 
radius of this cone increases with growing $\eta$.

\subsection{Perceptron learning with normalized weights}
\label{GSL-PN}
A similar calculation can be done for the perceptron
learning rule with normalized weights ($PN$), where
the length $\w{i}$ of the weight vectors is set to $1$ 
after each step. The perceptrons
move on a hypersphere of radius 1; in equilibrium, 
the average learning step leads back onto that sphere 
before the vectors are normalized again.

We derive the following differential equation for $R=\c$: 
\be
\dd{R}{\a} = (R+1)\left(\sqrt{\frac{2}{\pi}} \eta (1-R)
 - \eta^2 \frac{\theta}{\pi} \right). 
\ee
Fixed points are $R=1$, $R=-1$ and 
\be
\frac{\eta}{\sqrt{2 \pi}} \frac{\theta}{1-\c} =1. 
 \label{GSLPN-fix}
\ee
It is not a coincidence that this is equivalent
to (\ref{GSL-fix}) if $w$ is set to 1.
The fixed point of (\ref{GSLPN-fix}) at $R=1$ is 
repulsive; the one at 
$R=-1$ is unstable for $\eta < 4/\sqrt{2 \pi} \doteq 1.60$.
A solution of (\ref{GSLPN-fix}) can only be found for
$\eta \leq \eta_c \doteq 1.816$, which corresponds 
to $\c \doteq -0.689$.

\begin{figure} 
\epsfxsize= 0.65\textwidth
  \epsffile{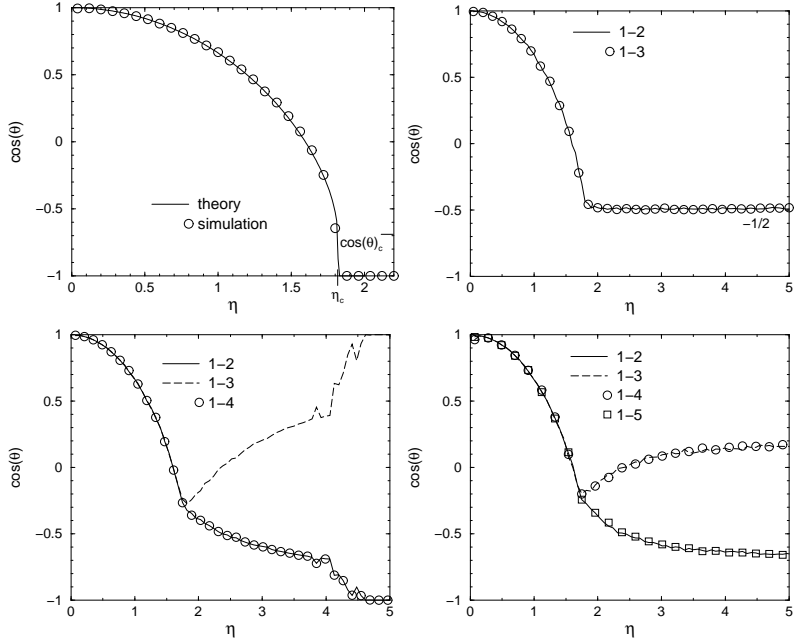}
\caption{Mutual learning with rule $PN$ (cf. sections 
\ref{GSL-PN} and \ref{GSL-K}): the system follows 
Eq. (\ref{GSLPN-fix}) for $\eta<\eta_c$.
Simulations used $N = 100$.
} 
\label{GSLN-fig}
\end{figure}

Simulations show that the system relaxes to the fixed point
given by Eq. (\ref{GSLPN-fix}) for $\eta < \eta_c$ and
jumps to $R = -1$ for larger $\eta$ (see Fig. \ref{GSLN-fig}).
This behaviour shows the 
characteristics of a first-order phase transition.

Hence for small learning rates the two perceptrons relax to 
a state of nearly complete agreement, $\theta \approx 0$.
Increasing $\eta$ leads to a nonzero angle between the 
two vectors up to $\theta \approx 133^{\circ}$. At this rate the 
system jumps to complete disagreement, $\theta = 180^{\circ}$. 

\subsection{Mutual learning on a ring}
\label{GSL-K}
The mutual learning-scenario can be generalized to
$K$ perceptrons: perceptron $i$ learns from perceptron 
$i+1$  if they disagree, with cyclic boundary
conditions. Under rule $P$, the total sum of vectors
is conserved again: as many perceptrons take a step
in one direction as in the opposite. 

Performing the necessary averages for the equations
of motion would involve Gaussian integrals over
$K-1$ correlated variables with $\Theta$-functions --
it is not clear to us whether this can be done 
analytically in general cases.
However, we find in simulations that the fixed point for
rule $P$ is completely symmetric: there is only
one angle $\theta$ between all pairs
of perceptrons. Assuming that relation (\ref{GSL-fix})
still holds, and using the conservation of $\sum \vw{i}$,
one can derive
\be
\frac{\eta}{\sqrt{2\pi}}\frac{\theta}{1-\c} = 
\frac{w_0}{\sqrt{1 + (K-1)\c}}.
\label{GSLK-fix}
\ee
The largest angle that the perceptrons can take
is $\c = -1/(K-1)$, corresponding to a $K$-cornered
hypertetrahedron. This happens when $|\sum \vw{i}|$
is negligible w.r.t $\w{i}$. Simulations confirm 
that (\ref{GSLK-fix}) holds, as can be seen in 
Fig. \ref{GSLP-fig}

Similar to the case of two networks, all perceptrons 
agree with each other for small learning rate $\eta \rightarrow 0$.
For larger rates the system relaxes to a state of high 
symmetry  where all mutual angles
between the $K$ weight vectors are identical, $\theta_{ij} = \theta$.
Note that the symmetry is higher than  the topology of the 
flow of information (the ring). For high rates $\eta \rightarrow \infty$
the system relaxes to a state of maximal disagreement, i.e. the
largest possible mutual angle $\theta$ that is still compatible
with a symmetric arrangement.

For rule $PN$, the sum of the weights is not 
preserved. The fixed point of the dynamics 
follows the curve for two normalized weights
described by Eq. (\ref{GSLPN-fix}) in a completely
symmetric configuration. When the hypertetrahedron angle
is reached and $\sum \vw{i}$ vanishes, the symmetry
is partly broken. There are now different angles to
nearest neighbours, next-nearest neighbours etc.,
so the angles split up into $(K-1)/2$ different 
branches for odd $K$ and $K/2-1$ for even $K$.
Note that the system still has the symmetry of the ring.

With odd $K$, increasing $\eta$ increases the 
angle between nearest neighbours, up to some 
limit value. This angle is not the maximum 
nearest-neighbour angle allowed for by the geometric 
constraints, but seems to decrease with increasing K.

In the case of even $K$, simulations show a second 
transition at some higher value of $\eta$, where the
vectors split into two antiparallel clusters,
thus maximizing the nearest-neighbour angle.
The learning rate at which this transition typically
appears during the run of the program increases
with $N$. The conclusion is that the antiparallel
fixed point is not stable in the $N \rightarrow \infty$
limit, but de facto stable in simulations because
the self-averaging property of the ODEs breaks down 
at this point.
 
One may ask which symmetries survive if the
perceptrons are allowed to have different
individual learning rates. A close look reveals that
for rule $P$, there is a more general conserved 
quantity: $\sum_i^K \vw{i}/\eta_i$. Simulations show that 
the angles $\theta_{ij}$ again relax to a completely 
symmetric configuration depending on the average 
$\eta$ and the initial value of the 
new conserved quantity, while the norms
$w_i$ are proportional to the respective
learning rates $\eta_i$. 
For rule $PN$, variations in the learning
rates not only lead to slightly different
curves for each of the angles with individually
different $\eta_c$, they also suppress the
transition to the antiparallel state that is 
observed for even $K$. 

\subsection{Hebbian learning}
The reason why $P$ and $PN$ lead to antiparallel
orientation of the weight vectors for larger
learning rates is that they
concentrate on cases where the networks disagree.
Algorithms that reinforce what both networks 
agree on are more successful, as can be seen 
for rule $H$ for two perceptrons.

The differential equations are
\bea
\dd{\w{i}}{\a} & = & \eta \sqrt{\frac{2}{\pi}} \c +
           \frac{\eta^2}{2 \w{i}};  \nonumber \\
\dd{R}{\a} & = & \eta  \sqrt{\frac{2}{\pi}} (\w{1}+\w{2}) + 
      \eta^2 (1 - \frac{2 \theta}{\pi}). \label{GSLH-ODE}
\eea 
This system has no common fixed point, which means that
the $\w{i}$ grow without bounds. The asymptotic behaviour
can be seen from the equation for $\c$.
Assuming that $\w{1} = \w{2} = \w{}$, we find
\be
\dd{\c}{\a} = \frac{\eta}{\w{}} \frac{4}{\sqrt{2\pi}} (1 - \c^2) + 
             \frac{\eta^2}{\w{}^2} \left 
         ( 1 - \c - \frac{2\theta}{\pi}\right ).
\ee
By taking $\w{} \approx \sqrt{2/\pi} \eta \a$, the ODE
leads to $1-\c \propto \a^{-4}$ for $\a \rightarrow \infty$.
This means that $\theta  \propto \a^{-2}$.

\begin{figure} 
\epsfxsize= 0.65\textwidth
  \epsffile{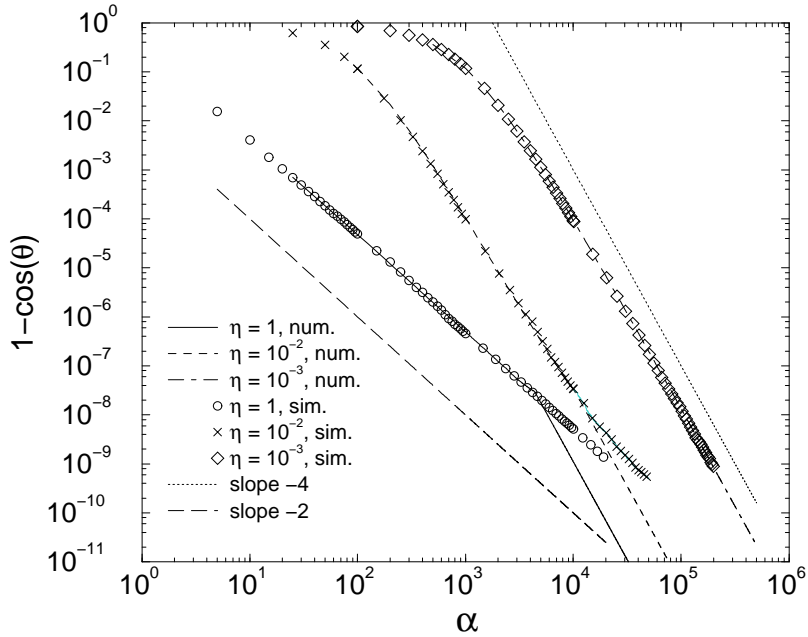}
\caption{Mutual learning with rule $H$: 
simulations with $N=100$ show  good agreement with Eqs. (\ref{GSLH-ODE}),
except for very small angles $\theta$.} 
\label{GSLH-fig}
\end{figure}

Simulations agree with the numerical integration of Eqs. 
(\ref{GSLH-ODE}), with the exception of very large $\a$ and
correspondingly small $\theta$ (see Fig. \ref{GSLH-fig}).
 This is not surprising, since
the $\a^{-2}$-decay is an effect of patterns that are
classified differently. As long as the perceptrons 
give the same output on all patterns, $\w{1}$ and $\w{2}$
grow linearly, but the difference $\vw{1}-\vw{2}$ does not
change, leading to $\theta\propto \a^{-1}$. This is 
observed in simulations for small angles, where no
patterns happened to be classified differently on the
considered timescale. Mathematically, this is related
to a breakdown of the self-averaging properties of 
Eqs. (\ref{GSLH-ODE}) at the point $\theta = 0$.

\section{Mutual learning, competition}
\label{SEC-Compet}
In the previous section, all of the neural networks 
behave in the same way. Each perceptron tries to learn
the output of its neighbour, and only the initial 
weight vectors are chosen randomly and differ from
each other. Now we investigate a scenario where two 
networks behave differently. Network 1 is trying to
simulate network 2 while 2 ist trained on the 
opposite of the opinion of 1. This scenario
describes a competition between two adaptive 
algorithms. If 2 is completely successful, the overlap is $\c=-1$,
and perceptron 1 always fails in its prediction, and vice versa.
A motivation from game theory can be drawn from the game 
of penny matching, where both players make a binary 
decision simultaneously. One player wins if the decisions
are the same, the other if they are different.

\subsection{Rule $P$}
If both perceptrons use rule $P$ for their respective 
learning aim, the update rules
are
\bea 
\vwp{1} &=& \vw{1} + (\eta_1/N) \x \sm{2} \Theta(- \sm{1} \sm{2}); 
 \nonumber \\
\vwp{2} &=& \vw{2} - (\eta_2/N) \x \sm{1} \Theta(\sm{1} \sm{2}).
\eea
The  corresponding differential equations for the order 
parameters are
\bea
\dd{\w{1}}{\a} &=& -\frac{\eta_1}{\sqrt{2\pi}}(1-\c) + 
     \frac{\eta_1^2}{2\w{1}}\frac{\theta}{\pi} \nonumber \\
\dd{\w{2}}{\a}&=& - \frac{\eta_2}{\sqrt{2\pi}}(1+\c)
+ \frac{\eta_2^2}{2 \w{2}}(1-\frac{\theta}{\pi}); \nonumber\\
\dd{R}{\a} &=& \frac{\eta_1 \w{2}}{\sqrt{2 \pi}} (1-\c) -
            \frac{ \eta_2 \w{1}}{\sqrt{2 \pi}}(1+\c).
\eea
The common fixed point for these equations is 
$\w{i}= \sqrt{2\pi}\eta_i/4$, $\c = 0$. This is hardly
surprising, since none of the perceptrons has a better
algorithm than the other. The learning rate only
rescales the weight vectors; the ratio $\eta_i/\w{i}$, 
which determines how fast the direction of $\vw{i}$ in
weight space can change, is independent of $\eta$ at 
the fixed point.

\subsection{Rule $H$}
The picture is slightly different if both perceptrons
learn from every pattern they see.
The resulting differential equations are
\bea
\dd{\w{1}}{\a} &=&  \sqrt{\frac{2}{\pi}} \eta_1 \c + 
     \frac{\eta_1^2}{2\w{1}} \label{KLZ-H-w1};\nonumber \\
\dd{\w{2}}{\a} & = & - \sqrt{\frac{2}{\pi}} \eta_2 \c + 
     \frac{\eta_2^2}{2\w{2}} \label{KLZ-H-w2};\nonumber  \\
\dd{R}{\a} & = &  \sqrt{\frac{2}{\pi}} \eta_1 \w{2}  - 
      \sqrt{\frac{2}{\pi}} \eta_2 \w{1} - \eta_1 \eta_2 (\pi - 2 \theta). 
\eea  
The fixed point of $R$ is reached if $\theta = \pi/2$ and 
$\eta_1/\w{1} = \eta_2/\w{2}$, i.e. the vectors are
perpendicular and the scaled learning rates $\eta_i/\w{i}$ are
the same for both perceptrons.
Under these conditions, the equations for $\w{i}$ can
be solved:  $\w{i} =\eta_i (\a +(\w{i,0}/\eta_1)^2)^{1/2}$,
so $\w{i}$ shows the $\sqrt{\a}$-scaling typical for 
random walks. 
Geometrically, the Hebb rule adds corrections
to the weight vector that are on average parallel to the
teacher vector. Since the teacher is moving at the same
angular velocity as the student, the movement of
both vectors resembles a random walk.
Again, $\eta$ only sets the temporal and
spatial scale.

\subsection{Rule $P$ vs. rule $H$}
The result of the competition becomes more 
interesting when both perceptrons use different
algorithms. For example, we let perceptron 1
use rule $P$, while 2 uses $H$. The derivation
of the differential equations is again straightforward:
\bea
\dd{\w{1}}{\a} & = & - \frac{\eta_1}{\sqrt{2\pi}}(1-\c)
     + \frac{\eta_1^2}{2 \w{1}} \frac{\theta}{\pi};\nonumber \\
\dd{\w{2}}{\a} & = & - \sqrt{\frac{2}{\pi}}\eta_2 \c 
          + \frac{\eta_2^2}{2 \w{2}};\nonumber \\
\dd{R}{\a} & = & - \sqrt{\frac{2}{\pi}} \eta_2 \w{1}
                 + \frac{\eta_1 \w{2}}{\sqrt{2\pi}}(1-\c)
                 + \frac{\eta_1 \eta_2 \theta}{\pi} 
                \label{KLZ-RBvsH}.
\eea
They have a common fixed point defined by
\bea
\theta \frac{\c^2}{(1-\c)^2} & = & \frac{\pi}{4}; \nonumber \\
\w{1} & = & \frac{\eta_1}{\sqrt{2\pi}} 
          \frac{\theta}{1-\c ;}\nonumber \\
\w{2} & = & \frac{\sqrt{2\pi}}{4}\frac{\eta_2}{\c}.
\eea
These equations can be solved numerically and yield
$\c \doteq 0.459$, $\w{1} \doteq 0.806 \eta_1$ and
$\w{2} \doteq 1.37 \eta_2$. Although perceptron 1
makes less use of the provided information, it
wins the competition: the perceptron using rule $H$
has a smaller $\eta/\w{}$-ratio and is thus less
flexible.

\subsection{Normalized weights}
By setting the weights to 1 after each learning step,
a new length scale is introduced, leading to a more complex
dependence of the solution on the learning rates.
For brevity, we only give the differential equations for the
different learning rules and explain some common features.
If both networks use rule $PN$, the ODE is
\be
\dd{R}{\a} = \frac{1}{\sqrt{2\pi}}(\eta_1(1-R) -\eta_2(1+R))
+ \frac{R}{\sqrt{2\pi}}(\eta_1(1-R) + \eta_2(1+R)) 
 - \frac{R}{2\pi}(\eta_1^2 \theta + \eta_2^2(\pi - \theta));
\label{KLZ-PNfix}
\ee
for rule $HN$ we find 
\be
\dd{R}{\alpha}=\sqrt{\frac{2}{\pi}} (\eta_2- \eta_1)(R^2-1)
-\frac{R}{2}(\eta_1^2 + \eta_2^2) - 
\eta_1 \eta_2 (1 - \frac{2\theta}{\pi}); \label{KLZ-HNfix}
\ee
and if rule $PN$ is used by perceptron 1 and $HN$ by 2, 
the equation is
\bea
\dd{R}{\alpha} &=& R\left (\sqrt{\frac{2}{\pi}}\eta_2 R - \frac{\eta_2^2}{2}
+ \frac{\eta_1}{\sqrt{2\pi}}(1-R) -
\frac{\eta_1^2}{2}\frac{\theta}{\pi}\right )\nonumber \\
& &  + \frac{\eta_1}{\sqrt{2\pi}}(1-R) - \sqrt{\frac{2}{\pi}}\eta_2
+ \eta_1 \eta_2 \frac{\theta}{\pi}. \label{KLZ-HNPNfix}
\eea

\begin{figure} 
\epsfxsize= 0.65\textwidth
  \epsffile{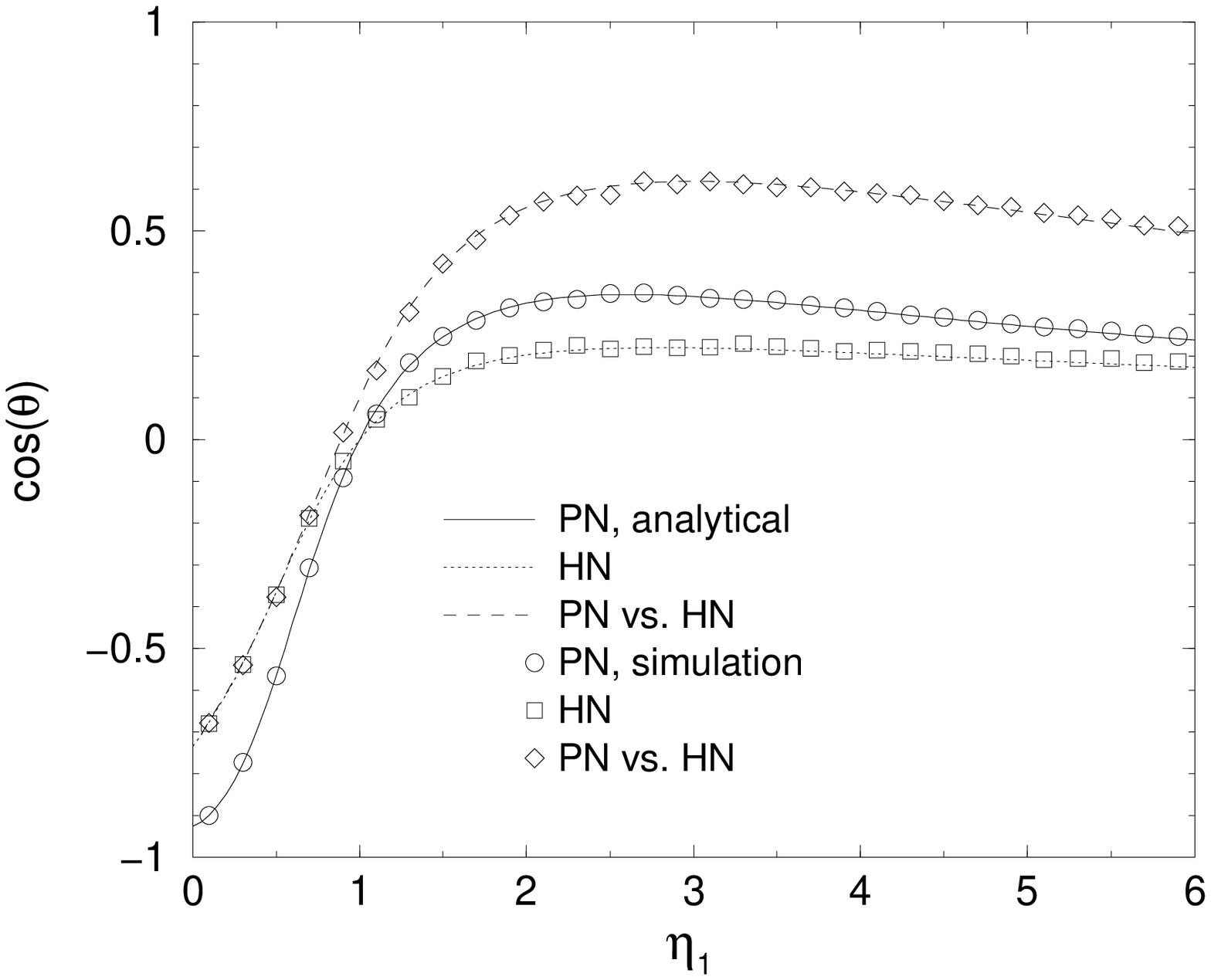}
\caption{Competing learning aims with normalized weights:
$\eta_2$ is set to 1 while $\eta_1$ is varied. 
The analytical curves are fixed points of Eqs. 
(\ref{KLZ-PNfix}), (\ref{KLZ-HNfix}) and (\ref{KLZ-HNPNfix}),
respectively.} 
\label{KLZ-XN-plot}
\end{figure}

The behaviour of the fixed point is similar in all cases
(see Fig. \ref{KLZ-XN-plot}):
\begin{itemize}
\item if, say, $\eta_2$ is fixed and $\eta_1 \rightarrow 0$,
$R$ goes to a value $R\neq-1$. This is expected, since
both $PN$ and $HN$ only achieve finite values of $R$ for
fixed teachers. 
\item if both perceptrons use the same algorithm with the 
same learning rate, the result is $R=0$, as expected.
\item if $\eta_i \rightarrow \infty$ for either $i$, 
$R \rightarrow 0$.
Infinite learning rate means that in every time step the
perceptron discards all the information it previously had,
replacing it with the current $\pm \x$. Theoretically,
that makes it predictable for the other network; in practice,
both agents are confused. The notable exception is the case
of $PN$ vs. $HN$, where a non-vanishing $R$ results if 
both $\eta_i \rightarrow \infty$ with a finite ratio
$\eta_1/\eta_2$.
\end{itemize} 
  

\section{Confused Teacher}
\label{SEC-Conteach}
For any prediction algorithm there is a bit sequence for which this
algorithm fails completely, with $100\%$ error \cite{Kinzel:Seq.}.
In fact, such a sequence is easily constructed: Just take 
the opposite of the predicted bit at each time step.
In Ref. \cite{Kinzel:Seq.} a perceptron was used for the prediction 
algorithm.

Here we do not consider bit sequences. However, it turns 
out that many statistical properties of the prediction
algorithm are similar when random inputs are used instead
of a window of the antipredictable bit sequence.
Hence  we consider the following scenario:
Preceptron 1 is trained on the negative of its own
output. Perceptron 2 is trained on the output of 
perceptron 1.  

This is similar to the teacher/student model where the 
teacher weight vector performs a random walk \cite{Biehl:Drift}.
But here the teacher is ``confused'', it does not believe its own
opinion and learns the opposite of it.

The update rule of perceptron 1 now only depends on its
own output:
\be
\vwp{1}  = \vw{1} - (\eta/N) \x \sm{1}.
\ee
Geometrically speaking, the vector performs a directed
random walk in which every learning step has a negative
overlap with the current vector. An equilibrium length
is reached when a typical learning step leads back
onto the surface of an $N$-dimensional hypersphere.
This fixed point of $\w{1}$ is easily calculated to be
\be
\w{1} = \sqrt{2 \pi} \eta/4 \doteq 0.6267 \eta, \label{VL-fix}
\ee
and the weight vector typically moves on the surface
of a hypersphere of that radius.

\subsection{Rule $H$}
What happens if a second perceptron tries to follow the
output of the confused teacher? Again, the results
depend entirely on the used algorithm. The simplest
case, the Hebb rule, also has a geometrical interpretation 
that is revealed  by a look at the update rule:
\bea
\vwp{1} & =& \vw{1} - (\eta/N) \x \sm{1}; \nonumber \\
\vwp{2} & = & \vw{2} + (\eta/N) \x \sm{1}.
\eea
As in section \ref{GSL-P}, the sum of both vectors is
constant, so there is a class of solutions to 
the ODEs
\bea
\dd{\w{1}}{\a} &=& -\sqrt{\frac{2}{\pi}} \eta + 
      \frac{\eta^2}{2 \w{1}}; \nonumber \\
\dd{\w{2}}{\a} &=& \sqrt{\frac{2}{\pi}} \eta \c + 
      \frac{\eta^2}{2 \w{2}}; \nonumber \\
\dd{R}{\a} &=& \sqrt{\frac{2}{\pi}}\eta (\w{1} - \w{2} \c) + \eta^2
\eea
defined by $\w{1,f} = \sqrt{2 \pi} \eta/4$ and 
$\w{2,f} = - \sqrt{2 \pi} \eta /(4\c )$. The solution is
given by the initial condition, i.e. the initial sum
$|\vw{1} + \vw{2}|$. The fixed point angle can be
calculated by applying the cosine theorem to a triangle
with side lengths $\w{1,f}$, $\w{2,f}$ and $|\vw{1} + \vw{2}|$;
starting from perpendicular vectors of norm $\w{0}$, one finds
\be
\c = - \left (1+ \frac{16}{\pi} 
       \left(\frac{\w{0}}{\eta} \right)^2 \right)^{-1/2}.
\ee
Geometrically, for large learning rate $\eta$ both norms become much 
larger than $\w{0}$; the only way to achieve this  while keeping
the sum constant is a large angle. For small $\eta$, 
$\w{1} $ becomes very small compared to the sum, and thus to
$\w{2}$. So the direction of $\vw{2}$ stays nearly unchanged while $\vw{1}$
performs its random walk, leading to nearly perpendicular
vectors on average.

\subsection{Rule $P$}
If perceptron 2 uses rule $P$, the sum of the vectors is
not conserved, and a simple geometrical interpretation
is not possible.
However, the equations of motion  can still be solved:
\bea 
\dd{\w{1}}{\a} &=& -\sqrt{\frac{2}{\pi}} \eta + 
      \frac{\eta^2}{2 \w{1}}; \nonumber \\
\dd{\w{2}}{\a} &= & - \frac{ \eta}{\sqrt{2 \pi}} (1-\c) + 
 \frac{\eta^2}{2 \w{2}} \frac{\theta}{\pi};\nonumber \\
\dd{R}{\a} &=& \sqrt{\frac{2}{\pi}} \eta \c -
   \frac{\w{1} \eta}{\sqrt{2\pi}} (1-c) - \eta^2 \frac{\theta}{\pi}.
\eea 
The fixed point of $\c$ is given by the solution of
$4 \theta/\pi  = (1+ \c)^2$, independent from $\eta$. The
numerical solution is $\theta \doteq 0.777 \pi$, $\c = -0.761$,
$\w{2} = 0.552 \eta$ (in accordance with Ref. \cite{Kinzel:Seq.},
where a special case of this problem was solved). Remarkably, the
generalization error is larger than 50\% - even the ``smarter''
perceptron learning rule predicts the behaviour of the 
confused teacher with less success than random guessing would.

\subsection{Optimal learning rule}
\begin{figure} 
  \epsfxsize= 0.65\textwidth
  \epsffile{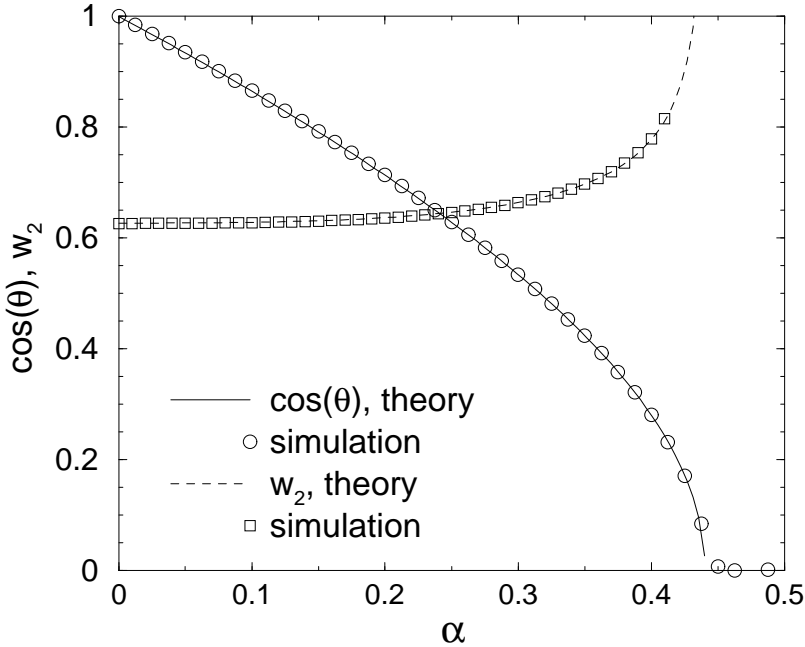}
  \caption{Confused teacher: Even with the optimal weight 
    function (\ref{VL-fopt}) the student only achieves an 
    overlap of $\c=0$. Starting values are $\w{1} = \w{2} = \sqrt{2\pi}/4$,
    $\c = 1$, $\eta =1$.
    Simulations are performed with $N=2000$; the statistical error is 
smaller than the size of the symbols..} 
  \label{VL-fopt-plot}
\end{figure}

This raises an interesting question: is there any 
``reasonable'' algorithm for perceptrons that allows
them to track the confused teacher? If there are
algorithms that achieve a positive overlap, one of them
has to be the rule that optimizes student-teacher
overlap in each time step -- the optimal weight function
derived by Kinouchi and Caticha \cite{Kinouchi:Optimal}:
\be
f_{\mbox{opt}} = \frac{\w{2} \tan (\theta)} {\sqrt{2 \pi}}
\exp \left [ -\frac{(\sp{\x}{\vw{2}})^2}
  {2\tan (\theta)^2\; \w{2}^2} \right ] 
\frac{1}{\Phi (\sm{1} \sp{\x}{\vw{2}}/(\w{2} \tan (\theta)))},
\label{VL-fopt}
\ee
where $\Phi(x) = \int_{-\infty}^{x} \exp(-z^2/2)/\sqrt{2 \pi} \d{z}$.
If $\w{1}$ is set to its fixed point for simplicity's sake,
calculation yields the following ODEs for $\c$ and $\w{2}$:
\bea
\dd{\c}{\a} &=& \frac{1}{4 \pi} \frac{\sin (\theta)^2}{\c}\, I 
    -\frac{2}{\sqrt{2 \pi} \w{1} \c}; \label{VL-fopt-eq1}\\
\dd{\w{2}}{\a} &=& \frac{\w{2}}{4 \pi} \tan (\theta)^2\, I, 
 \mbox{\ \ \  where} \label{VL-fopt-eq2}\\
I &=& \int_{-\infty}^{\infty} \frac{1}{\sqrt{2 \pi}} 
 \exp\left(- \frac{1+ \c^2}{2\sin(\theta)^2} x^2 \right)
 \frac{1}{\Phi(-x \cot (\theta))\Phi(x \cot(\theta))} \d{x}.
\eea 
Calculating whether $\c = 0$ is in fact a fixed point
of the confused teacher/optimal student scenario is
problematic, since the optimal weight function  
(\ref{VL-fopt}) diverges at $\theta = \pi/2$. However,
the numerical solution of Eqs. (\ref{VL-fopt-eq1}) and 
(\ref{VL-fopt-eq2}) shows clearly that even starting from 
$\c =1$, the system evolves  towards 
$\theta = \pi/2$, which indeed seems to be the upper limit
for success. Simulations of the learning
process again agree weel with our theory (see Fig. \ref{VL-fopt-plot}).

\subsection{Rule $HN$}  
\begin{figure} 
  \epsfxsize= 0.65\textwidth
  \epsffile{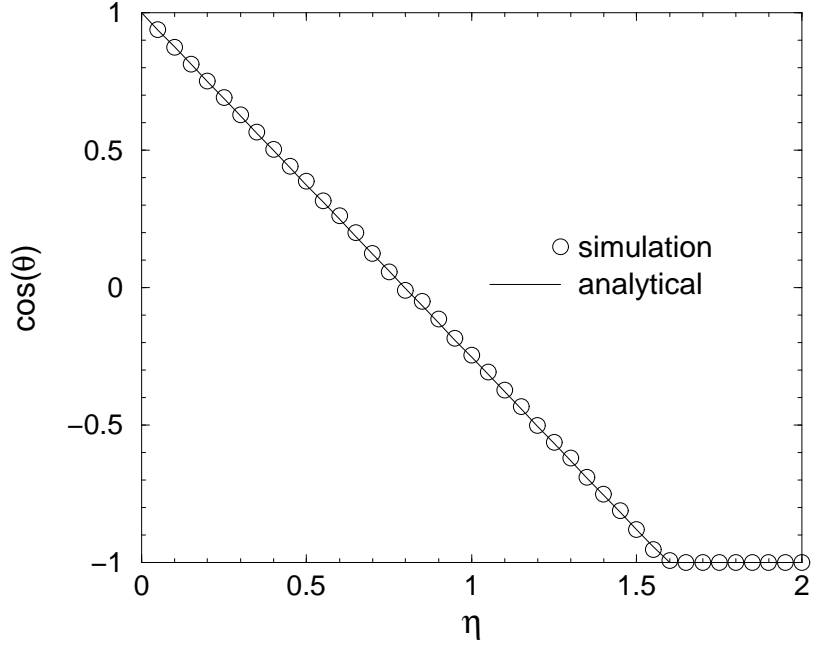}
  \caption{Confused teacher: If the teacher is slowed down by
    normalizing its weight, it can be tracked by a student using
    e.g. rule $HN$. The figure shows the fixed point of (\ref{VL-HNfix})
    and simulations with $N=100$.} 
  \label{VL-HN-fig}
\end{figure}

There is a way of achieving a positive overlap with the
confused teacher with simple learning rules: if the
teacher perceptron is ``slowed down'' by keeping its weights 
normalized and setting $\eta$ to some small value, 
a student using $PN$ or $HN$ can track the
teacher nearly perfectly for very small learning rates. 
For simplicity's sake, let us consider $HN$ with identical 
learning rates. The differential equation for $R$ is
\be
\dd{R}{\a} = (R+1) 
  \left ( \sqrt{\frac{2}{\pi}}(1 -R)\eta -\eta^2 \right ), \label{VL-HNfix}
\ee
the fixed points are $R = -1$ or $R = -\sqrt{2/\pi} \eta +1$.
This result is again confirmed by simulations, as seen in Fig. 
\ref{VL-HN-fig}. The fixed point goes to $1$ as $\eta \rightarrow
0$.

\section{Perceptrons in the Minority Problem}
\label{SEC-Minority}
The concept of  interacting neural networks can be applied to 
a problem that has received much attention recently:
the El Farol Bar Problem \cite{Arthur:ElFarol}.
The problem was originally inspired by a popular bar
that has a limited capacity: if too many people
attend, it becomes crowded, and patrons don't enjoy
the evening. In a more special formulation, each 
agent out of a population of $K$ decides in each
time step (each Saturday evening) to take one
of two alternatives (go to the bar or stay at home).
Those agents who are in the minority win, the
others lose.
Decisions are made independently; the only information
available to agents is the decision of the minority was in the
last $N$ time steps. 

Many papers (see e.g. \cite{Minority}) investigated a specific realisation
of the model called the Minority Game. 
In this model each agent has a small number of randomly chosen
decision tables (Boolean functions)
that prescribe an action based on the previous history,
and which of the tables is used is decided according to how 
successful each one was in the course of the game. 
It turned out that the success of the game depends
on the ratio between the number of players and the 
size of the history window, and
general conclusions on the behaviour of crowded
markets were drawn \cite{Challet:Modeling,Johnson:Trader}.

We will discuss a different approach that yields different 
behaviour: Each agent $i$ is represented by a 
perceptron $\vw{i}$ that uses the time series $\v{S}_t
=(S_t, S_{t-1}, \ldots,S_{t-N+1})$
of past minority decisions to make a prediction on
the next time step. 
It then learns the output of the minority according
to some learning rule. 

In our approach, all of the agents are flexible in their decisions.
Each agent uses an identical adaptive algorithm which is trained
by the history of the game, the only information available
to each of the agents. However, each agent uses a different
randomly chosen initial state of its network.
If all weight vectors of the networks would collapse, all
agents would make the same decision, and all would lose.
If all weights remained in the random initial state,
each agent would make a random guess which yields a 
reasonable performance of the system. Our calculation
shows that training can improve the performance of the
system compared to the random state.

Following Ref. \cite{Cavagna:Memory}, we replace the
history $\v{S}_t$ by a random vector $\x$. Simulations
show that this changes the results only quantitatively,
if at all. 

This strategy fulfills the restrictions that 
the original problem posed: the agents do 
not communicate except through majority 
decisions, and individual decisions are based on experience
(induction or learning) rather than perfect knowledge 
of the system (deduction). However, since each player
uses only one strategy whose parameters can be fine-tuned 
to the current environment rather than a set of completely
different strategies, no quenched bias in the players' 
behaviour is to be expected.   

\subsection{General notes on performance}
The commonly used measure of collaboration in the
minority problem is the average standard deviation 
of the sum of outputs of all agents:
\be
\frac{\sigma^2}{K} = \frac{1}{K} \langle (\sum_{i=1}^{K} \sm{i})^2 \rangle.
\label{MIN-sm1}
\ee
If each agent makes random decisions, one gets $\sigma^2/K = 1$.
The probability of two perceptrons $i$ and $j$ giving the same 
output on a random pattern is $1-\theta_{ij}/\pi$.
Any ensemble of vectors $\vw{i}$ can be thought of as
centered around a center of mass $\C = \sum_{i=1}^{K} \vw{i}/K$
with a norm $C$ (for random vectors of length 1, $C$ would
be of order $\sqrt{K}$).
The weights can then be written as $\vw{i} = \vg{i} +\C$,
with $\sum_{i=1}^{K}  \vg{i} =\v{0}$. For the sake of 
simplicity, we will assume a symmetrical configuration
with $g_i = 1$ and $\sp{\vg{i}}{\vg{j}} = -1/(K-1)$ for $i\neq j$.
(An ensemble of randomly chosen vectors of norm 1 would
give $g_i^2 = 1-1/K \pm O(1/\sqrt{N})$ and
$\sp{\vg{i}}{\vg{j}} = -1/K \pm O(1/\sqrt{N})$.)

The average overlap between different weights is now 
$R = C^2 - 1/(K-1)$, their average norm $w_i = \sqrt{\C^2+1}$.
With this, Eq. (\ref{MIN-sm1}) can be evaluated:
\bea 
\frac{\sigma^2}{K}  &=& \frac{1}{K} \left \langle \sum_{i=1}^{K} 1 +
\sum_{i=1}^{K} \sum_{j\neq i}^{K} \s{\sp{\x}{\vw{i}}} \s{\sp{\x}{\vw{j}}}
\right \rangle_{\x} \nonumber \\
&=& 1 + (K-1) \left (1 - \frac{2}{\pi} 
\arccos \left (\frac{C^2 - 1/(K-1)}{C^2+1}
\right ) \right ). \label{MIN-sm2}
\eea
If $C$ is set to 0 and $K$ is large, a linear expansion of 
the arccos term in Eq. (\ref{MIN-sm2}) gives $\sigma^2_{opt}/K 
\approx 1- 2/\pi \doteq 0.363$. The small anticorrelations
(of order $1/K$) between the vectors suffice to change the 
prefactor in the standard deviation.

If $C$ is much larger than $g$, there is a strong correlation
between the perceptrons. Most perceptrons will agree with
the classification by the center of mass $\s{\sp{\x}{\C}}$.
As $C \rightarrow \infty$, $\sigma^2/K$ saturates at $K$.

\subsection{Hebbian Learning}
Now each perceptron is trying to learn the decision of 
the minority according to rule $H$. $S$ denotes 
the majority decision:
\be 
\vwp{i} = \vw{i} - \frac{\eta}{M} \x\, \s{\sum_{j=1}^{N}  \s{\sp{\x}{\vw{j}}}}
  =  \vw{i} - \frac{\eta}{M} \x\, S.
\ee
As the same correction is added to each weight vector, their mutual
distances remain unchanged. Only the center of mass is shifted.
We now treat $C$ as an order parameter:
\bea
\C^+ &=& \sum_{i=1}^{K}\frac{ \vw{i}}{N} - 
       \frac{\eta}{M} \x\, s. \\
{C^2}^+ &=& C^2 -\frac{2 \eta}{N} \sp{\x}{\C} S + \frac{\eta^2}{N}.
\eea
To average over $\sp{\x}{\C} S$ in the thermodynamic limit,
we introduce a  field $h = \sp{\x}{\C}$ and average over $\x$ for
fixed $h$:
\be
\sp{\x}{\C} S = |h| \s{\sum_{i=1}^{K} \s{h}\s{\sp{\x}{\vg{i}}+h}}.
\label{MIN-xCS}
\ee
The quantity $\s{h} \s{\sp{\x}{\vg{i}}+h}$ is a random variable with mean 
$\erf(|h|/\sqrt{2})$ and variance $1- \erf(|h|/\sqrt{2})^2$.
In a linear approximation for small $|h|$, we replace this by
mean $\sqrt{2/\pi}|h|$ and variance 1.

\begin{figure} 
  \epsfxsize= 0.65\textwidth
  \epsffile{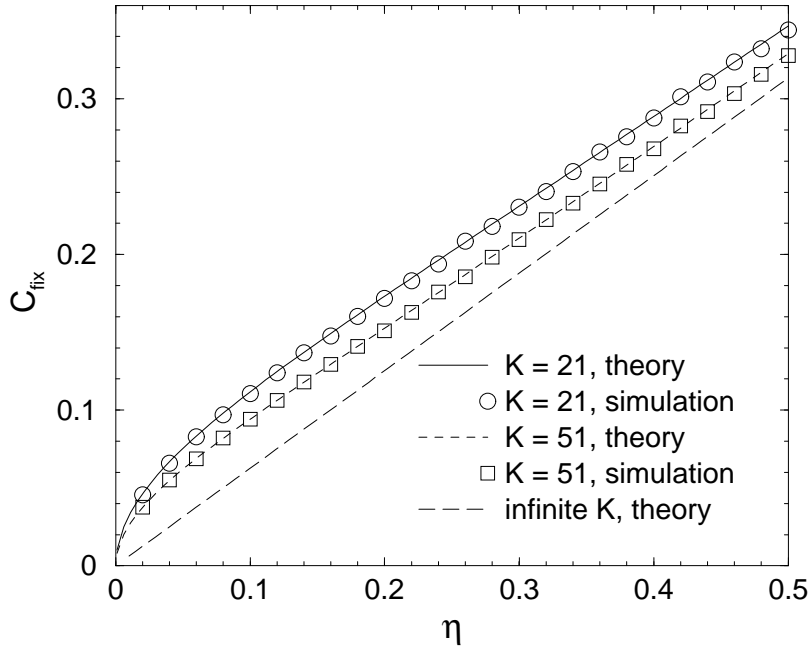}
  \caption{Fixed point of $C$ vs. $\eta$: simulations with $N=100$
    agree well with Eq. (\ref{MIN-Cfix}). The limit for $K \rightarrow
    \infty$ is $C = \sqrt{2 \pi} \eta/4$. } 
  \label{MIN-C_eta}
\end{figure}

\begin{figure} 
\epsfxsize= 0.65\textwidth
  \epsffile{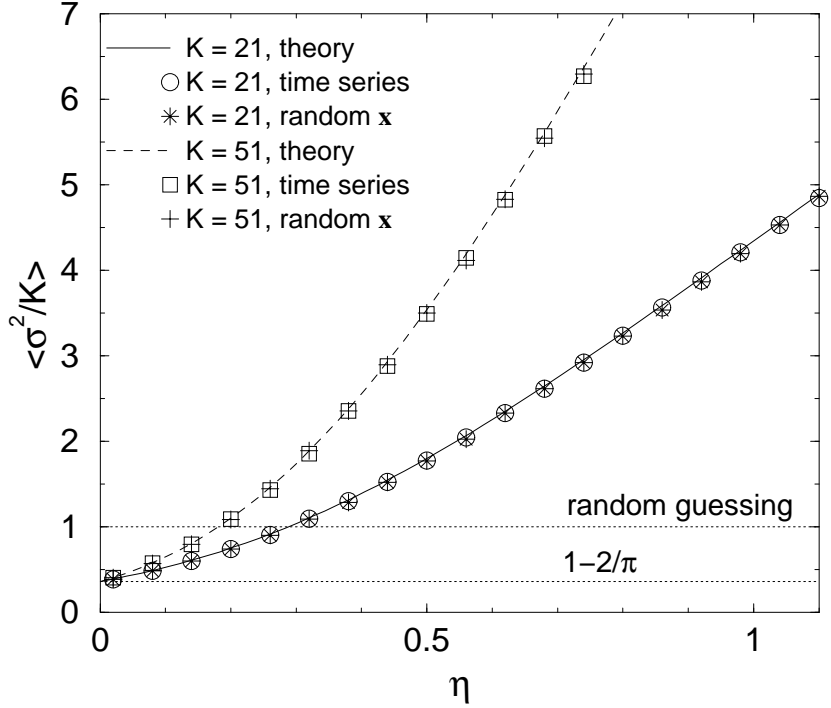}
  \caption{Fixed point of $\sigma^2/K$ vs. $\eta$: the
    combination of Eqs.(\ref{MIN-sm2}) and (\ref{MIN-Cfix})
    shows that sufficiently small learning rates lead to
    $\sigma^2/K <1$.} 
  \label{MIN-sm_eta}
\end{figure}

For sufficiently large $K$, one can use the Central Limit Theorem 
to show that $\sum_{i=1}^{K} \s{h}\s{\sp{\x}{\vg{i}}+h}$
becomes a Gaussian random variable with mean $\sqrt{2/\pi} K |h|$.
Since the terms of the sum in (\ref{MIN-xCS}) 
are anticorrelated rather than independent,
the variance turns out to be $(1-2/\pi) K$ rather than $K$, 
analogously to Eq. (\ref{MIN-sm2}). This yields
\be
\left \langle \s{\sum_{i=1}^{K} \s{h} \, \s{\sp{\x}{\vg{i}} + h}}
 \right  \rangle = 
\erf (\sqrt{K/(\pi-2)}|h| ). \label{MIN-indep} 
\ee
Since $h$ is a Gaussian variable with mean $0$ and variance $C^2$,
the average over $h S$ can now be evaluated. We find the 
following differential equation for the norm of the center 
of mass:
\be 
\dd{C^2}{\a} = -\frac{4 \eta}{\sqrt{2 \pi}} 
    \sqrt{\frac{2 K/(\pi -2)}{1 + 2 K(\pi -2)C^2}} C^2 +\eta^2.
\ee
The fixed point of $C$, which can be plugged into Eq.  
(\ref{MIN-sm2}) to get $\sigma^2/K(\eta,K)$, is 
\be
C = \frac{\sqrt{\pi}}{4} \eta \sqrt{1 +\sqrt{1+ 
  \frac{16 (\pi-2)}{\pi K \eta^2}}} \label{MIN-Cfix}
\ee
(see Figs. \ref{MIN-C_eta} and \ref{MIN-sm_eta}).

If $C$ is large, 
the majority of perceptrons will usually make
the same decision as $\C$, which then behaves like the single confused
perceptron: $C \rightarrow \sqrt{2 \pi} \eta/4$ if 
$K \eta^2 \rightarrow \infty$ -- compare to Eq. (\ref{VL-fix}).

For small $C$, the majority may not coincide with $\s{\sp{\x}{\C}}$.
In that case, the learning step has a positive overlap 
with $\C$, leading to $C \propto \sqrt{\eta}$ as $\eta \rightarrow 0$.

The derivation given is only correct if $N \rightarrow \infty$
and $K$ is large. However, simulations show very good agreement even
for $K =21$ and $N=100$ (see Fig. \ref{MIN-sm_eta}). 
For a smaller number of dimensions
$N$, there is even a tendency towards smaller $\sigma^2/K$.
This can be understood in the extreme case of $N=1$:
Each perceptron is characterized by one number; the outcome
is decided by whether the majority of numbers is smaller
than $0$ or larger, regardless of the ``pattern''. 
The learning step consists of shifting all numbers up or
down by the same amount. In the case of small 
$\eta$, the fixed point is characterized by $(N-1)/2$
players firmly on one side of the origin, $(N-1)/2$
on the other side, and one unfortunate loser who changes
sides at every step.

Interestingly, if the time series generated by the
minority decisions is used as patterns, the
functions $\sigma^{2}(C)$ and $C(\eta)$
are quantitatively different from those found
for random patterns. However, in the final result 
$\sigma^2(\eta)$ no disagreement can be noticed 
(see Fig. \ref{MIN-sm_eta}).

\begin{figure} 
  \epsfxsize= 0.65\textwidth
  \epsffile{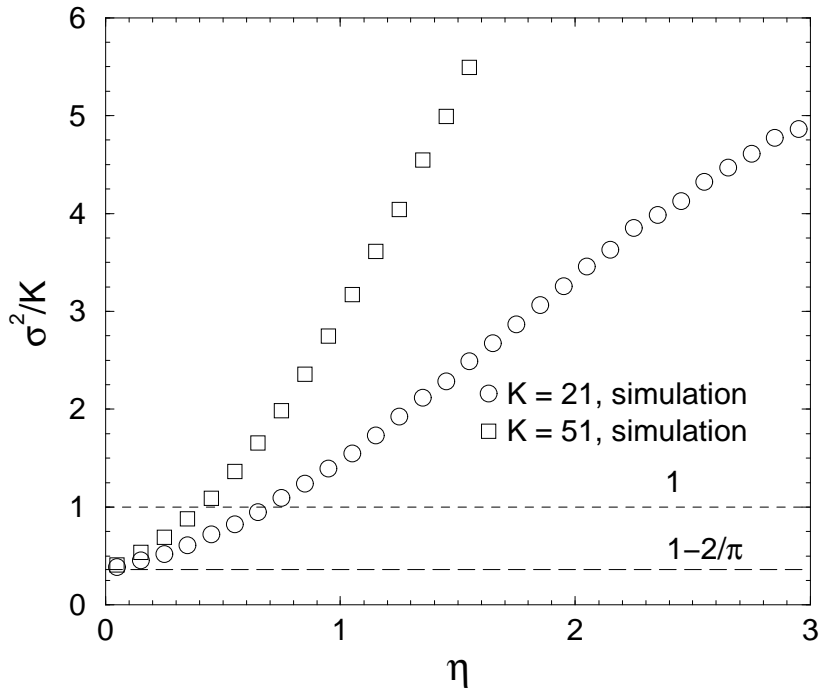}
  \caption{Using a modified $PN$ algorithm improves the 
    results, compared to Fig. \ref{MIN-sm_eta}. Simulations 
    again use $N=100$.} 
  \label{MIN-invPerz}
\end{figure}

The presented Hebb algorithm may appear too simplistic
and the chosen initial conditions too artificial.
It must therefore be emphasized that there are other
learning algorithms that lead to the same anticorrelated
state. In particular, a variation of rule $PN$ has proven
successful in simulations (see Fig. \ref{MIN-invPerz}):
all perceptrons that are on the minority side take a 
learning step, and weights are kept normalized.
The regular rule $P$ where perceptrons on the majority side
move, however, leads to strong clustering and 
$\sigma^2/K \propto K$.

The absence of scaling behaviour if $N > K$ and the fact
that smaller dimensions (corresponding to smaller memory
of the time series) even improve the results show that
the conclusions drawn from the ``conventional'' Minority
Game do not apply to all conceivable strategies for
the Bar Problem. We think that the dependence of $\sigma^2/K$
on the ratio between available strategies and players
is caused by the use of quenched strategies and will
not arise in any scenario in which agents stick to
one strategy which is fine-tuned by some learning
process.

The case of $N=1$ implies
that there are strategies that give $\sigma^2/K \propto 1/K$.
We will elaborate this point in another publication.

\section{Summary}
We have investigated several scenarios of mutually 
interacting neural networks. Using perceptrons with
well-known on-line training algorithms in the limit
of infinite system size, we derived exact equations
of motion for the dynamics of order parameters which
describe the properties of the system.
In the first scenario a system of $K$ perceptrons 
is placed on a ring.
 All perceptrons receive 
the same input and each perceptron is trained by the
output of its neighbour on the ring.
We have used two well-known training algorithms:
the perceptron rule which concentrates on examples
where the networks disagree, and the Hebbian rule 
where each example changes the weights. We find
that with unnormalized weights the system relaxes 
to a stationary state of high symmetry: each perceptron
has the same overlap with all others. The overlap depends
on the learning rate: with increasing $\eta$ the perceptrons
increase their mutual angle as much as possible.

For the perceptron learning rule with normalized weights we find
phase transitions with increasing learning rate $\eta$.
For large values of $\eta$, the symmetry is broken, but 
the symmetry of the ring is still conserved.
For the Hebbian rule we find a different behaviour. 
The lengths of the weights diverge, the mutual angles
shrink to zero and the perceptrons eventually come to
perfect agreement in the limit of infinitely many training
examples.
  
We furthermore study the behaviour of perceptrons
that pursue competing learning aims for different
learning algorithms. If two perceptrons follow
mutually exclusive learning aims using the same
algorithm, a draw results. If they use different
rules, the outcome depends on factors like the
rescaled learning rate $\eta/w$.  
We find that a perceptron that learns the opposite
of its own prediction cannot be tracked by a student
perceptron that learns the positive output of the 
confused teacher: all rules achieve a negative overlap.

Finally an ensemble of interacting 
perceptrons is used to solve a model of a closed
market. Each agent uses a perceptron which is trained 
on the decision of the minority.  Our analytic solution shows that
the system relaxes to a stationary state which yields
a good performance of the system for small learning rates
$\eta$. In contrast to the minority game of Refs. 
\cite{Minority} our approach leads to identical profits for all agents
in the long run. In addition, the performance of the
algorithm is insensitive to the size of the history 
window used for the decision.

This paper is a first step towards more complex models
of interacting neural networks. We have presented 
analytically accessible cases which may open 
the road to a general understanding of interacting adaptive
systems with possible applications in biology, computer
science and economics.

\section{Acknowledgement}
All authors are grateful for support by the GIF. 
This paper also benefitted from a seminar at the 
Max-Planck-Institut komplexer Systeme, Dresden.
We thank Johannes Berg, Michael Biehl,  Liat Ein-Dor,
Andreas Engel, Georg Reents, and Robert Urbanczik 
for helpful discussions. 

\section{Appendix}
\label{Appendix}
The following averages are used in our calculations
to derive deterministic differential equations from
the update rules. The angled brackets denote 
averages over isotropically distributed pattern
vectors. In the limit $N\rightarrow \infty$, 
$\sp{\vw{1}}{\x}$ and $\sp{\vw{2}}{\x}$ are 
correlated gaussian random variables, and the
averages can be calculated by integrating over
their joint probability distribution with appropriate
boundaries. In many cases, simple geometrical 
calculations give the same result with less effort.

\bea 
\left \langle \sp{\x}{\vw{1}}\, \sm{2} 
   \Theta(-\sm{1} \sm{2}) \right \rangle &=&
         - \frac{\w{1}}{\sqrt{2 \pi}}(1 -\c);  \\
\left \langle \sp{\x}{\x}\,  
   \Theta(-\sm{1} \sm{2}) \right \rangle &=&
         N \frac{\theta}{\pi};  \\
\left \langle \sp{\x}{\vw{1}}\, \sm{1}  
   \Theta(\sm{1} \sm{2}) \right \rangle &=&
         \frac{\w{1}}{\sqrt{2 \pi}}(1 +\c);  \\
\left \langle \sp{\x}{\x}\,  
   \Theta(\sm{1} \sm{2}) \right \rangle &=&
         N (1- \frac{\theta}{\pi}); \\
\left \langle \sp{\x}{\vw{1}}\, \sm{1}  \right \rangle &=&
    \sqrt{\frac{2}{\pi}} \w{1};\\
\left \langle \sp{\x}{\vw{1}}\, \sm{2}  \right \rangle &=&
    \sqrt{\frac{2}{\pi}} \w{1} \c;\\
\left \langle f_{\mbox{opt}} \right \rangle &=&
        \frac{2 \w{2}}{\sqrt{2 \pi}} \frac{\sin(\theta)^2}{\c};\\
\left \langle f_{\mbox{opt}}\, \sp{\x}{\vw{2}}\, \sm{1} \right \rangle &=&
     0; \\
I &=& \int_{-\infty}^{\infty} 
   \frac{1}{\sqrt{2 \pi}} 
 \exp \left (- \frac{1 + \c^2}{2 \sin(\theta)^2} x^2 \right )
\left (\Phi ( - x \cot (\theta)) 
        \Phi ( x \cot (\theta)) \right )^{-1} \d{x};\\ 
\left \langle f_{\mbox{opt}}^2 \right \rangle &=&
        \frac{\w{2}^2}{2 \pi} \tan (\theta)^2\, I;  \\
\left  \langle f_{\mbox{opt}}\, \sp{\x}{\vw{1}}\,\sm{1} \right \rangle &=&  
        \frac{\w{1} \w{2}}{2 \pi} \frac{\sin(\theta)^2}{\c}\, I.
\eea

\end{document}